\RequirePackage{fix-cm}
\documentclass[smallextended]{svjour3}       
\smartqed  
\usepackage{graphicx}
\journalname{Foundations of Physics}

\begin{document}

\title{Are Quantum-Classical Hybrids compatible with Ontological Cellular Automata?}

\titlerunning{Are Quantum-Classical Hybrids compatible with Ontological Cellular Automata?}        

\author{Hans-Thomas Elze}

\institute{Hans-Thomas Elze \at 
Dipartimento di Fisica ``Enrico Fermi'' \\ Universit\`a di Pisa, Largo Pontecorvo 3, I-56127 Pisa, Italia \\  
\email{elze@df.unipi.it}           
}

\date{Received: date / Accepted: date}

\maketitle

\begin{abstract} 
Based on the concept of ontological states and their 
dynamical evolution by permutations, as assumed in the Cellular Automaton Interpretation (CAI) of quantum mechanics, we address the issue whether quantum-classical hybrids can be described consistently in this framework.  
We consider chains of `classical' two-state Ising spins and their discrete deterministic dynamics 
as an ontological model with an unitary evolution 
operator generated by pair exchange interactions. 
A simple error mechanism is identified, which turns them into quantum mechanical objects, chains of qubits. Consequently, an interaction between a quantum mechanical and a `classical' chain 
can be introduced and its consequences for this   quantum-classical hybrid be studied. We find that 
such hybrid character of composites, generally, does not persist under interactions and, therefore, cannot be upheld consistently, or even as a fundamental notion {\it \`a la} Kopenhagen interpretation, within CAI.  
\end{abstract}

\section{Introduction} 
Our aim presently is to apply recent results illustrating the Cellular Automaton Interpretation of quantum mechanics  
\cite{tHooft2014,tHooft2020,tHooft2021,ElzeQu19,ElzePAFT19,Elze2020} and reconsider the status of Quantum-Classical 
Hybrids \cite{Elzehybrid}, in particular. Can such hybrids exist and consistently be described in 
a theoretical framework according to this interpretation? We summarize necessary ingredients here, as well as in the following sections, which will ultimately lead to a negative answer to this question.  

\subsection{Quantum-Classical Hybrids -- a reminder}
Quantum-Classical Hybrids (QCH) have been of interest since the early days when quantum theory has obtained its canonical formulation, as represented in well known textbooks \cite{Dirac,vonNeumann}. 

In the beginning, to separate a composite system into 
a {\it classical} part and a {\it quantum mechanical} part, as far as their respective states and 
dynamical behaviour are concerned, may have been mostly of practical interest as an approximation method for otherwise intractable situations \cite{BornOppenheimer} -- which, to this day,  continues to be essential in quantum chemical studies of complex molecules, even biomolecules, and their reactions; similarly, for example, in nuclear reactions, {\it etc.}  

However, eventually this topic also gained attention from the perspective of foundational issues in quantum theory \cite{Sudarshan}. In particular, numerous and varied attempts to understand and resolve 
the infamous {\it measurement problem} of quantum mechanics (QM) have continued to 
flourish, which we shall not review here.\,\footnote{However, for a review and a recent study, see the 
Refs.\,\cite{Nieuwenhuizen,Konishi}.} -- they all, in one way or another, try to deal with the separation of measurement readings of a 
classical apparatus, {\it i.e.}, corresponding to non-superposed pointer states, from the quantum evolution of the object under study. No consensus has been reached, how to resolve this conundrum {\it within} quantum theory -- yet the Cellular Automaton Interpretation of QM does offer an elegant resolution, to be mentioned in Section\,1.2.     

Lately, questions of the existence (or not) of QCH have newly moved into focus, since the fundamentally important debate about the either classical or quantum mechanical nature of gravitation is soon to be accompanied by (possibly table-top) high precision experiments, see, 
{\it e.g.}, \cite{LampoEtAl2014,Hu,Ulbricht2017,Ulbricht2021,MarlettoVedral}. 

Presently, we are interested to consider the consistency or inconsistency of QCH on the 
basis of the Cellular Automaton Interpretation of QM. In the past, there have been numerous attempts to provide a theoretically satisfactory description of QCH, trying to embed the formalisms of QM and classical mechanics into a common framework, especially along the lines of Ref.\,\cite{Heslot} (with references to earlier work there). However, {\it interacting} QCH 
present numerous features, where consistency can fail. An almost complete list of essential  
consistency checks has been presented and passed in Ref.\,\cite{Elzehybrid}. Additional issues, however,  have been discussed in Refs.\,\cite{Salcedo,Elze2012,Diosi2012,Diosi2014}, for example.  

\subsection{Cellular Automaton Interpretation of QM in a nutshell}  
According to the Cellular Automaton Interpretation (CAI) of QM, the evolution of the 
Universe is deterministic and happens in discrete steps \cite{tHooft2014}. 

{\it Ontological States} ($\cal OS$) are the discrete physical states the Universe can be in. At each step of its evolution, 
the Universe is in a definite state, which we may denote, for example, by   
$$
|A\rangle ,|B\rangle ,|C\rangle ,\dots \;; 
$$ 
for simplicity, we assume a finite (or countably infinite) number $N$ of states. 
 
{\it No superpositions} of the $\cal OS$ exist physically ``out there''.  
It is only by {\it permutations} that $\cal OS$ evolve into $\cal OS$, {\it e.g.}, 
$$  
|A\rangle\rightarrow |B\rangle\rightarrow |C\rangle\rightarrow\dots \;. 
$$ 
		
Thus, the set of $\cal OS$ forms a {\it preferred basis}. 
We may declare this basis to be orthonormal and define an associated {\it Hilbert space}. 

Diagonal operators on the $\cal OS$ are {\it beables} and their eigenvalues correspond to the 
labels $A, B, C,\dots$ by which we labeled the ontological states. 

Mathematically, by unitary transformations of the preferred basis we arrive at 
{\it Quantum States} $\cal QS$ as superpositions of $\cal OS$. These constructs serve to do physics 
with the help of mathematical language, according to CAI. 

An important property of $\cal QS$ follows 
immediately from the evolution of $\cal OS$ by permutations. Referring to the above example, a  
generic quantum state,  
\begin{equation}\label{QuSt} 
|Q\rangle :=\alpha |A\rangle +\beta |B\rangle +\dots\;,
\;\; |\alpha |^2+|\beta |^2 +\dots\; =1 
\;\;, \end{equation}  
evolves accordingly:  
\begin{equation}\label{OntCons}
|Q\rangle\; \rightarrow\; \alpha |B\rangle +\beta |C\rangle +\dots
\;\;, \end{equation} 
where the amplitudes $\alpha$, $\beta$, {\it etc.} are conserved. This conservation law has been termed the 
{\it conservation of ontology}. 

If we {\it choose} the amplitudes introduced in the $\cal QS$ of Eq.\,(\ref{QuSt})   
to encode probabilities $|\alpha |^2$, $|\beta |^2$, {\it etc.} of the $\cal OS$  
$|A\rangle$, $|B\rangle$, {\it etc.}, respectively, to present the initial state of 
the evolution step (\ref{OntCons}), then we have the {\it Born rule} of QM 
built right into the formalism. Notably, the {\it conservation of ontology} under  
evolution is an essential ingredient.   

Completing this sketch of CAI, {\it Classical States} ($\cal CS$) are considered as probabilistic 
distributions of $\cal OS$. Typically, they describe physical macrosystems, which cannot be 
represented in terms of individual $\cal OS$. 

Experiments that ``happen'' repeatedly -- like apparently repeating evolutions of a sufficiently but always incompletely isolated part of the Universe -- pick up different initial conditions regarding $\cal OS$. 
Therefore, the {\it classical apparatus} in an experiment must generally be expected to yield 
different pointer positions as outcomes. Due to the {\it conservation of ontology}, the  
probability of a particular outcome directly reflects the probability of having a particular $\cal OS$ as 
initial condition, since it evolves by permutations of elements of this preferred basis only, 
{\it cf.} (\ref{OntCons}).

Thus, according to CAI, the apparent {\it reduction or collapse} to a single pointer position as a measurement result 
arises due to the intermediary use of {\it quantum mechanical} (superposition) {\it states}, when describing what in reality 
are evolving $\cal OS$ that differ in different runs of an experiment. 

Summarizing, the Cellular Automaton Interpretation suggests to build 
{\it discrete deterministic dynamical models} with evolution generated by 
{\it permutations of ontological states}. 
The periodicity of trajectories in the space of states, {\it i.e.} without fusion or 
fission, turns them into QM models, as has been demonstrated in several studies 
before, see, {\it e.g.}, 
Refs.\,\cite{tHooft2014,tHooft2020,tHooft2021,ElzeSchipper,ElzeRelativPart,PRA2014,Margolus}.\,\footnote{We mention that 
various demonstrations exist of how to arrive at QM with   
statistical arguments or with stochastic or dissipative modifications of {\it classical deterministic} dynamics, 
{\it i.e.} without postulating an underlying ontology, such as 
\cite{HabaKleinert,Khrennikov2005,BlasoneEtAl,SakellariadouEtAl,IsidroEtAl,DArianoEtAl,Wetterich2010,Wetterich2021}.}
 
We have seen how from rather parsimonious and simple assumptions 
about the underlying ontology one arrives at essential features of QM. While other aspects will be 
discussed in the following, we conclude here with a few remarks concerning open problems of CAI 
that need to be addressed in the future. 

First of all, it turns out to be difficult to incorporate 
{\it interactions} into previously noninteracting models. This problem, however, cannot be sidestepped, 
if anything like an interacting quantum field theory, especially the paradigmatic Standard Model, is 
to be explained by an ontological underpinning along the lines of CAI. 
Furthermore, the relation of the dynamics generated by permutations of ontological states to 
a {\it Hamiltonian} that determines the unitary evolution operator as in QM is generally not 
straightforward to work out. Some glimpses of this will be encountered in what follows.   
 
In Sections\,2 and 3, we mainly reformulate results obtained in Refs.\,\cite{ElzeQu19,ElzePAFT19,Elze2020} in such a way that our argument against the consistency of quantum-classical hybrids, according to 
CAI, follows easily, as given in Section\,4. 
Conclusions are presented in Section\,5. 

\section{Permutations of ontological states} 
Let $N$ objects, $A_1,A_2,\dots ,A_N$ (``states''), be mapped in $N$ steps onto one another, 
involving {\it all} states. 
Suitably arranging the sequence of states, this may always be represented by an 
{\it unitary} $N\times N$ matrix with {\it one} off-diagonal phase per column and row and zero  elsewhere:     
\begin{equation}\label{Umatrix}
\hat U_N:= 
\left (
\begin{array}{c c c c c} 
0           & .           & .     & 0       & e^{i\phi_N} \\ e^{i\phi_1} &0            &       & .       & 0  \\ 
0           & e^{i\phi_2} & .     &         & .  \\
.           &             & .     & .       & .  \\
0           & .           & 0     & \;e^{i\phi_{N-1}} & 0  \\
\end{array}\right )
\;\;. \end{equation} 
We associate with this particular representation a basis of vectors called the 
{\it standard basis}. 

It is easy to see that the matrix $\hat U_N$ has the significant property: 
\begin{equation}\label{UmatrixN} 
(\hat U_N)^N=e^{i\sum_{k=1}^N \phi_k}\;\mathbf{1}
\;\;. \end{equation} 
This implies that the {\it Hamiltonian}, defined by the 
relation $\hat U_N=:e^{-i\hat H_NT}$, can be immediately diagonalized 
to give the diagonal matrix elements ($n=1,...,N$): 
\begin{equation}\label{Hdiag}	
(\hat H_N)_{nn}=\mbox{diag}\Big (\frac{1}{NT}\big (2\pi (n-1)
        -\sum_{k=1}^N\phi_k\big )\Big ) 
\;\;, \end{equation} 
which refer to the {\it diagonal basis}. In the following, the arbitrary phase angles $\phi_k$ play 
no role, so we set them to zero.  

Naturally, the diagonal and standard bases are unitarily related. The 
transformation between the bases can be obtained explicitly by a 
discrete Fourier transformation \cite{Elze2020}. 
Using this result, in turn, one evaluates the Hamiltonian for the standard basis, 
to find the diagonal and off-diagonal matrix elements, respectively: 	
\begin{equation}\label{Hstandard1}
(\hat H_N)_{nn}=\frac{\pi}{NT}(N-1)\;,\; n=1,\dots ,N 
\;\;, \end{equation} 
and, 
\begin{equation}\label{Hstandard2} 
(\hat H_N)_{n\neq m}=\frac{\pi}{NT}
		\Big (-1+i\cot\big (\frac{\pi}{N}(n-m)\big ) \Big ) 
		\;,\;\; n,m=1,\dots ,N  
\;\;, \end{equation} 
and $\hat H_N=\hat H_N^\dagger$. 
	
\begin{figure}  
\begin{center}
\includegraphics[width=0.65\columnwidth 
]{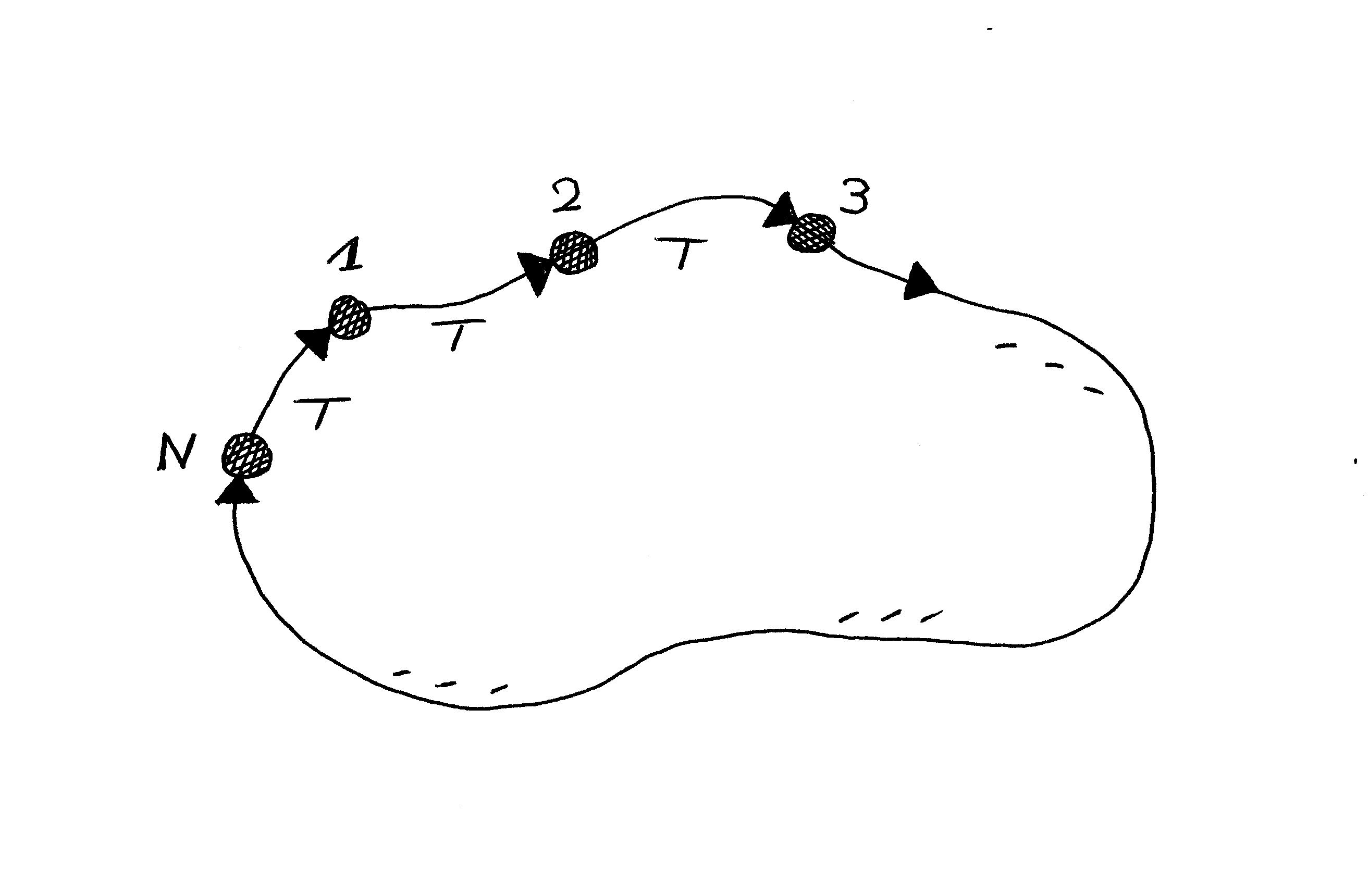}   
\end{center} 
\caption{The cogwheel model of a system jumping periodically through $N$ states, with $T$ denoting the time it takes per jump.} 
\label{fig:1} 
\end{figure}

A {\it cogwheel model} describes the discrete deterministic dynamics 
by a unitary permutation matrix $\hat U_N$ \cite{tHooft2014}. A sketch 
is shown in Fig.\,\ref{fig:1}. 

It may come as a surprise, however, in the limits 
$N\rightarrow\infty $, $T\rightarrow 0$, with $\omega :=1/NT$ fixed, 
the simple cogwheel model describes the {\it quantum harmonic oscillator} 
or systems that can be related to it 
\cite{tHooft2014,ElzeSchipper,ElzeRelativPart}. Cellular automata which lead to more general 
quantum models in the continuum limit are studied in 
Ref.\,\cite{PRA2014}. 

In the next sections we will apply the dynamics of cogwheel models 
to study the behaviour of particular Ising spin systems. We choose them as  
classical discrete deterministic many-body systems evolving by permutations of 
their states, in line with the picture of ontological cellular automata.  

\section{Ising spin chains} 
Let a onedimensional spin chain with periodic boundary condition be  
composed of $2S+1$ {\it classical two-state spins}, which represent $2^{2S}$ $\cal OS$. 
We denote the spins and their values by, respectively: 
\begin{equation}\label{IsingSpinVal}
s_k=\pm 1=\;\uparrow ,\downarrow\; ,\;k=1,\dots ,2S+1,\;s_{2S+1}\equiv s_1 
\;\;, \end{equation} 
where the first and last spin are identified.   

Permutations of the $\cal OS$ of the chain will be generated by  		
{\it spin exchange} transpositions acting on pairs of spins:   
\begin{equation}\label{transpos} 
\hat P_{ij}|s_i,s_j\rangle :=|s_j,s_i\rangle \;, \;\;  
\hat P_{2S\;2S+1}\equiv\hat P_{2S\;1} 
\;\;, \end{equation} 
with $\hat P_{ij}\equiv\hat P_{ji}$; 
from here on we shall often use the {\it ket} notation familiar from QM 
to write down states of two or more Ising spins. These transpositions 
have the properties 
$\hat P^2=\mathbf{1}$ and $[\hat P_{ij},\hat P_{jk}]\neq 0$, for $i,j,k$ different from each other.  

Furthermore, it is well known but crucial fact that spin exchange operators or transpositions can 
be expressed in terms of a vector $\underline{\hat\sigma}$ formed by the three Pauli matrices 
$\hat\sigma^{x,y,z}$: 
\begin{equation}\label{PijPauli} 
\hat P_{ij}= (\underline{\hat\sigma}_i\cdot\underline{\hat\sigma}_j+\mathbf{1})/2 
\;\;. \end{equation}   

Finally, we define the {\it dynamics} of the spin chain by a particular choice of the 
unitary matrix $\hat U$ incorporating the transpositions that produce permutations 
of the $\cal OS$: 
\begin{equation}\label{dynamics} 
\hat U:=\prod_{k=1}^S\hat P_{2k-1\;2k}\prod_{l=1}^{S}\hat P_{2l\;2l+1}
	=:\exp (-i\hat HT) 
\;\;, \end{equation} 
with the associated Hamiltonian $\hat H$ to be extracted next.  
This evolution operator respects  
several conservation laws, among them translation invariance, as discussed in Ref.\,\cite{Elze2020}.  

We note that $\hat U$ distinguishes {\it even/odd pairs} $ij$, according to whether $i$ of $\hat P_{i<j}$ is even/odd. Following Eq.\,(\ref{dynamics}), all even pairs are updated first, which is followed by the update of all odd pairs, when $\hat U$ is applied to a given state once. All even pair transpositions 
$\hat P_{2l\;2l+1}$ commute with each other but not with the odd ones and {\it vice versa}. Therefore,  
the effect of certain transposition is felt by neighbouring pairs only when $\hat U$ is applied a second 
time. This implements a {\it final signal velocity} for the propagation of perturbations along the chain. 

\subsection{Extracting the spin-exchange Hamiltonian} 
Consider a generic initial $\cal OS$ of a chain 
of $2S$ Ising spins, 
\begin{equation}\label{initialchain}    
|\psi\rangle :=|s_1,s_2,s_3,s_4,s_5,\dots ,s_{2S-1},s_{2S}\rangle 
\;\;, \end{equation} 
where the periodic boundary condition means that  
$s_1\equiv s_{2S+1}$; such that the rightmost transposition in $\hat U$ 
of Eq.\,(\ref{dynamics}), $P_{2s\;2s+1}$ can meaningfully 
be applied to the state. 

\begin{figure} 
\begin{center} 
\includegraphics[width=0.9\columnwidth 
]{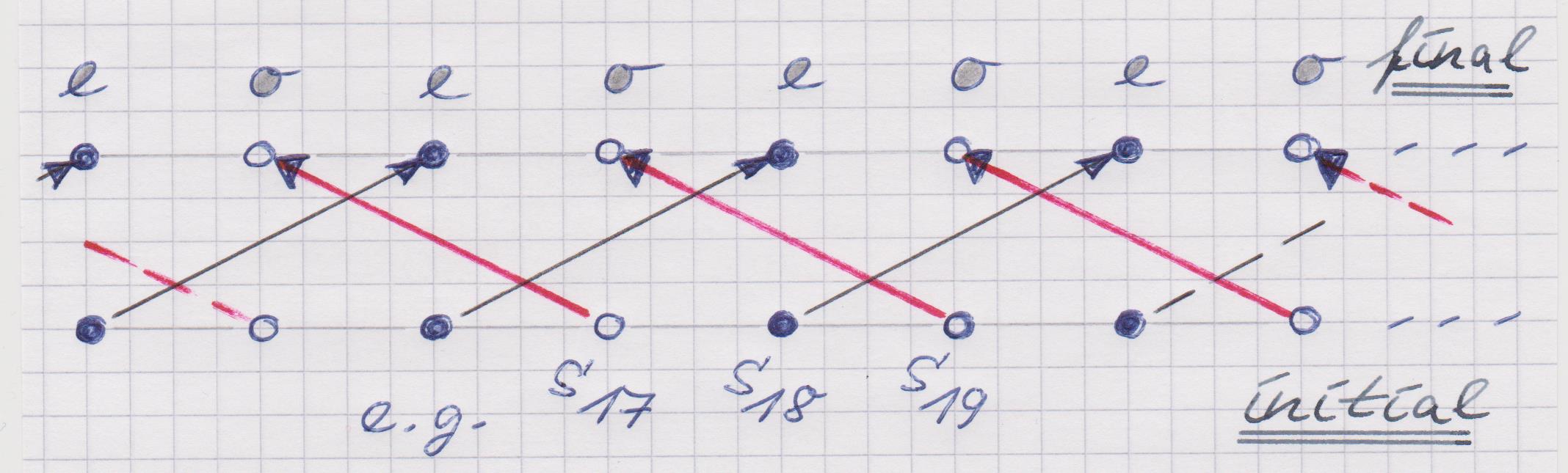}  
\end{center} 
\caption{One-step update by $\hat U$, Eq.\,(\ref{dynamics}), of a piece of an initial spin chain state (bottom), {\it cf.} Eq.\,(\ref{initialchain}). The state of a spin on 
site $k$, $s_k$, jumps two sites to the left (right) if $k$ is {\bf o}dd ({\bf e}ven), which produces the final state (top).}  
\label{fig:2} 
\end{figure}

Applying $\hat U$ once, 
we observe that the resulting evolution of the state is very simple, 
which reflects the conservation laws present. This is illustrated in 
Fig.\,\ref{fig:2}. Consecutive updates lead to repetitions of this pattern 
of {\it leftmovers} and {\it rightmovers}. 

If we follow the motion of a particular left- or   rightmover, it will consecutively pass all odd or even sites, respectively, and continue along the chain periodically. 
Thus, we find in this model a 
composition of cogwheels, {\it cf.}  
Section\,2, which is produced by 
nearest-neighbour transpositions in the classical spin chain. 
Since after $S$ updates the initial $\cal OS$ of the chain is always recovered, the latter itself behaves like an $S$-state cogwheel. 

In Ref.\,\cite{Elze2020}, we indicated how this 
motion can be encoded in two discrete {\it two-component 
field equations}. Which, 
in turn, can be mapped one-to-one on (1+1)-dimensional  
partial differential equations for bandwidth-limited 
classical fields with the help of {\it Sampling Theory}, {\it e.g.} similarly as in Ref.\,\cite{PRA2014}. 
	
We now briefly outline how the 
Hamiltonian $\hat H$ is extracted, once we have  
recognized that any $\cal OS$ of the chain 
evolves like a state in a $S$-state cogwheel, however,    
with an internal structure of $2S$  
Ising spins and nearest-neighbour exchange 
interactions. 

Since the overall behaviour is that of a $S${\it -state cogwheel}, the Hamiltonian must be related to 
$\hat H_{N=S}$, {\it cf.} Eqs.\,(\ref{Hstandard1})--(\ref{Hstandard2}). 
Following simple examples treated explicitly in 
Refs.\,\cite{ElzeQu19,ElzePAFT19}, it was shown in Ref.\,\cite{Elze2020} how to `dress' $\hat H_{N=S}$, in order to obtain $\hat H$. Namely, the latter has to act on the spin variables of a generic state $|\psi\rangle$, Eq.\,(\ref{initialchain}), in one-to-one correspondence to the former. This results in: 
\begin{eqnarray}\label{H10} 
\hat H&=&\sum_{n=1}^S(\hat H_S)_{n1}\hat U^{n-1} 
\;\; \\ [1ex] \label{H30}  
&=&\frac{\pi}{T}\Big (\mathbf{1}
+\frac{i}{S}\sum_{n=1}^{S-1}\cot (\frac{\pi}{S}n )\hat U^{n}\Big )
\;\;,\;\mbox{for}\;\;\hat U|\psi\rangle\neq |\psi\rangle 
\;\; \\ [1ex] \label{H40} 
&=&\frac{\pi}{T}\Big (\mathbf{1}
+\frac{i}{2S}\sum_{n=1}^{S-1}\cot (\frac{\pi}{S}n )
\big (\hat U^{n}-(\hat U^\dagger )^{n}\big )\Big )
\;\;. \end{eqnarray} 
Instead, for states with $\hat U|\psi_0\rangle =|\psi_0\rangle$, one consistently finds that $\hat H|\psi_0\rangle =0$ for these static {\it zero modes} --  
examples of these have all spins either up or down or all leftmovers up (down) and all rightmovers down (up). 
Taking into account that $\hat U^{S-k}=(\hat U^\dagger )^k$, which follows 
from $\hat U^k\hat U^{S-k}=\mathbf{1}=\hat U^k(\hat U^\dagger )^k$, for $1\leq k\leq S$, and the symmetry 
of the cot-function, the final result is given in a 
manifestly self-adjoint form.  

\subsection{Remarks}
In Ref.\,\cite{Elze2020} several comments were made about the interpretation of the Hamiltonian 
$\hat H$ that acts on $\cal OS$ of the Ising spin 
chain -- this concerns physical properties (degeneracy and magnetization of states, translation invariance), relation to a discrete field theory, and that formally the results summarized in this section yield an interesting, especially terminating {\it Baker-Campbell-Hausdorff formula}:   
\begin{equation}\label{BakerCH}
\hat{U}=
\exp \Big (-i\pi  
\big (\mathbf{1}
+\frac{i}{S}\sum_{n=1}^{S-1}\cot (\frac{\pi}{S}n )\hat U^{n}\big )\Big )
\;\;,\;\mbox{for}\;\;\hat U|\psi\rangle\neq |\psi\rangle 
\;\;, \end{equation}
with $\hat U$ of Eq.(\ref{dynamics}).  

We emphasize that the unitary operator $\hat U$, Eq.\,(\ref{dynamics}), with $\hat H$ of 
Eqs.\,(\ref{H10})--(\ref{H40}), has been constructed 
to evolve in a discrete, deterministic way states of  the {\it classical Ising spin model with exchange  interactions}. 
Most importantly, these $\cal OS$ evolve by permutations (transpositions) of the spin variables {\it without 
ever forming superposition states}, in accordance 
with CAI ({\it cf.} Section\,1.2).   

However, if numerical constants of the Hamiltonian, especially 
in Eq.\,(\ref{H40}), are only slightly perturbed, 
this becomes a genuine {\it quantum mechanical operator}. Generally, this will produce superposition 
states of $\cal OS$, which would necessitate a 
corresponding Hilbert space of $2S$ 2-state quantum spins for 
its description, {\it i.e.} QM of {\it qubits ``by mistake''} \cite{Elze2020}. -- Such perturbations could, for example, be caused by neglected or integrated out {\it fast high-energy degrees of freedom}, as suggested on other grounds recently  \cite{tHooft2021}.  

In any case, small errors in $\hat H$, or $\hat U$, tend to have a QM effect. For illustration, consider 
a long chain, with $S\gg 1$, and approximate $\hat H$, Eq.\,(\ref{H40}), 
roughly by the leading terms: 
\begin{equation}\label{H40approx}
\hat H\approx\frac{\pi}{T}\Big (\mathbf{1}
+\frac{i}{\pi}\big (\hat U-\hat U^\dagger\big )\Big )
\;\;. \end{equation}  
Then, choosing a state,   
$|\psi_{\downarrow\downarrow}\rangle :=|\;\dots\;,\uparrow ,\uparrow ,\uparrow ,\downarrow^e ,
\downarrow^o ,\uparrow ,\uparrow ,\uparrow ,\;\dots\;\rangle 
\;\;, $ 
where the two indicated down spins are located on neighbouring {\bf e}ven and {\bf o}dd sites, we obtain by definition of $\hat U$, Eq.\,(\ref{dynamics}): 
\begin{eqnarray}\label{Upsi} 
\big (\hat U-\hat U^\dagger \big  )|\psi_{\downarrow\downarrow}\rangle 
&=&\;\;\;|\;\dots\;,\uparrow ,\uparrow ,\downarrow ,\uparrow^e , \uparrow^o ,\downarrow ,\uparrow ,\uparrow  ,\;\dots\;\rangle 
\nonumber \\ [1ex] \label{Ustate} 
&\;&-|\;\dots\;,\uparrow ,\downarrow ,\uparrow ,\uparrow^e , \uparrow^o ,\uparrow ,\downarrow ,\uparrow  ,\;\dots\;\rangle  
\;\;, \end{eqnarray} 
and correspondingly $\hat H|\psi_{\downarrow\downarrow}\rangle$, by Eq.\,(\ref{H40approx}). 
This reminds of an {\it entangled Bell state}; it is not  
an $\cal OS$ but a quantum mechanical superposition state instead. 		

\section{Inconsistency of Quantum-Classical Hybrids in CAI}
Next, we employ our observations, in order to 
study the (in)consistency of Quantum-Classical Hybrids (QCH), {\it cf.} Section\,1.1, when confronted with CAI by way of a simple but representative example. 

We consider {\it two} Ising spin chains for this 
purpose, one which has become {\it quantum mechanical} by introducing superpositions of its $\cal OS$, {\it e.g.} by a mechanism similar to the one discussed 
above, and one which is {\it classical} in the form of a probability distribution of $\cal OS$, {\it cf.} Section\,1.2. 
In the simplest case, the latter may be sharply 
peaked on a single $\cal OS$.    

\begin{figure}[t] 
	\begin{center} 	
		\includegraphics[width=0.65\columnwidth 
		]{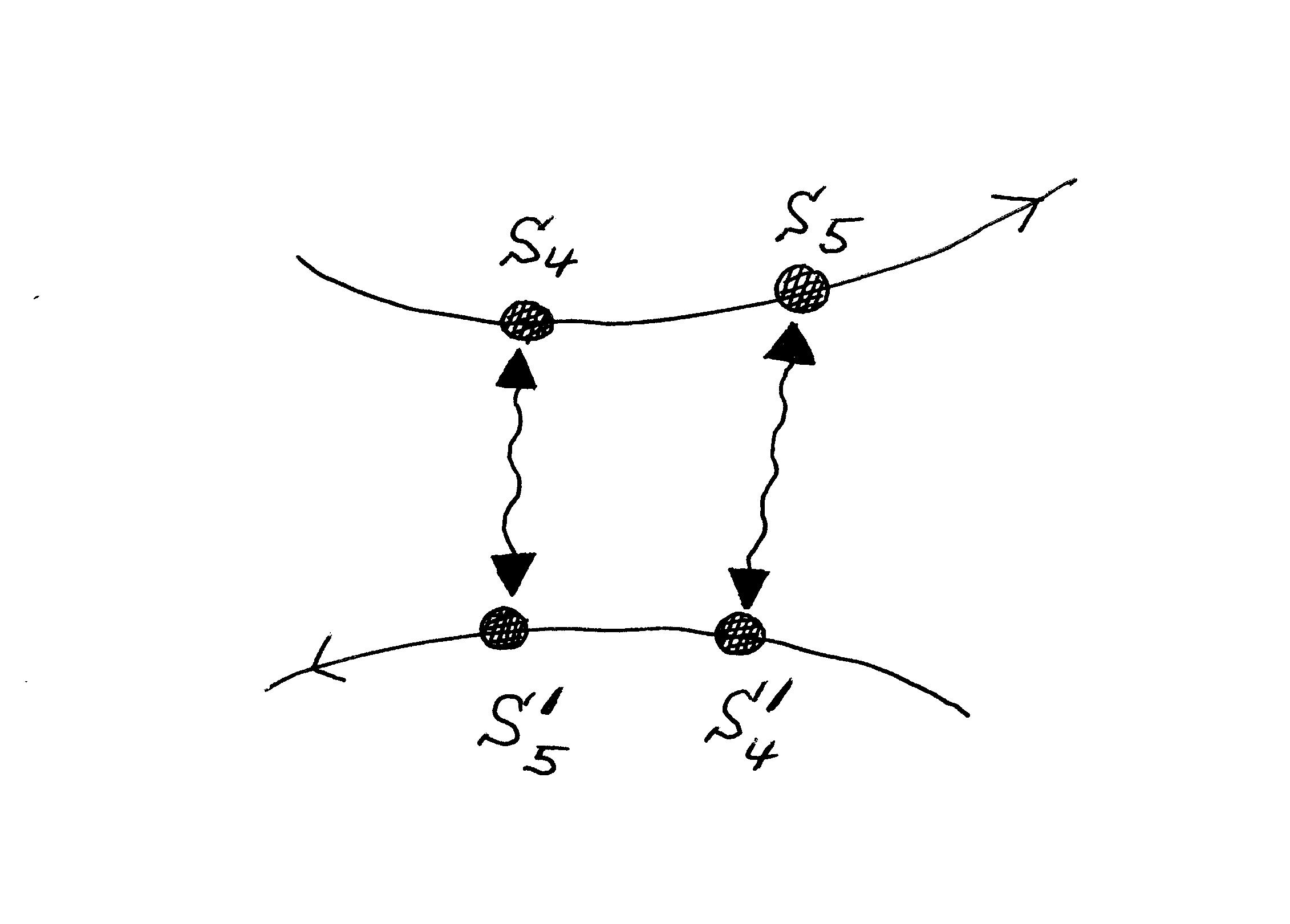} 
	\end{center}   
	\caption{The Ising spins at sites 4 and 5 of two chains, with spins labeled 
		by $s_k$ and $s'_k$, respectively, interact by exchanging spins $s_4$ and $s'_5$ and spins $s'_4$ and $s_5$. Thus, the states of a pair of {\it rightmovers} are exchanged with those of a pair of {\it leftmovers} by this momentum conserving interaction, {\it cf.} Section\,3.1.}  
	\label{fig:3} 
\end{figure}

Assuming a temporary interaction, which acts in between two particular updates of both chains, it should be of the ontological kind, {\it i.e.} 
involve only permutations (transpositions) of the Ising spins of the two chains. To be definite, we choose the exchange interaction 
illustrated in Fig.\,\ref{fig:3}. 

Let the initial {\it quantum state} of one of the two chains be given by the superposition: 
\begin{equation}\label{Qstate} 
|Q\rangle 
=\alpha |\dots ,a_3,a_4,a_5,a_6,\dots\rangle 
 +\beta |\dots ,b_3,b_4,b_5,b_6,\dots\rangle  
\;\;, \end{equation} 
with $|\alpha |^2+|\beta |^2=1$ and $a_k,b_k$ denoting specific values of the respective Ising spins. Furthermore, let the initial {\it classical state} of the 
second chain be given sharply by an $\cal OS$: 
\begin{equation}\label{Cstate} 
|C\rangle 
=|\dots ,s'_3,s'_4,s'_5,s'_6,\dots\rangle 
\;\;. \end{equation}   
Applying the interaction $\hat {\cal I}$ depicted in Fig.\,\ref{fig:3}, the QCH consisting of both chains is tranformed into: 
\begin{eqnarray}
\hat {\cal I}|Q\rangle|C\rangle 
&=&\;\;\alpha |\dots ,a_3,s'_5,s'_4,a_6,\dots\rangle
|\dots ,s'_3,a_5,a_4,s'_6,\dots\rangle
\nonumber \\ [1ex] \label{interaction} 
&\;&+\beta |\dots ,b_3,s'_5,s'_4,b_6,\dots\rangle
|\dots ,s'_3,b_5,b_4,s'_6,\dots\rangle  
\;. \end{eqnarray}   
Thus, such an interaction is a bilinear map  
$\hat {\cal I}:\{{\cal OS}\}\times\{{\cal OS}\}' \rightarrow {\cal T}$ on the direct product of spaces spanned by the preferred bases of ontological states of both chains. However, due to the linearity, linear superpositions of ${\cal OS}$ from the quantum state of one of the chains, {\it e.g.} Eq.\,(\ref{Qstate}), are 
transferred by the interaction into the target space 
${\cal T}$, which becomes the {\it Hilbert space}  
generated by the tensor product of those bases, 
${\cal T}=\{{\cal OS}\}\otimes\{{\cal OS}\}'$.   

A curious case arises, if both states in the 
superposition (\ref{Qstate}) differ only in one (or two) $a,b$ pair(s) out of ${a_4,a_5,b_4,b_5}$, but are 
identical otherwise: then, the  
feature of being in a quantum superposition state and 
in a classical state ($\cal OS$), respectively, is swapped between both chains. 

However, generally, the result of Eq.\,(\ref{interaction}) presents  an {\it entangled quantum state} involving 
both chains. In particular, the classical state $|C\rangle$ of the second chain is changed in such a way that it cannot be factored out anymore.   

Thus, the interaction changes here the character of the composite of two 
Ising chains from being a {\it quantum-classical hybrid} to an overall {\it quantum state}. This result follows similarly, if the quantum state in Eq.\,(\ref{Qstate}) is replaced by a more complicated superposition or the sharp classical state of the second chain is more 
generally replaced by a probabilistic mixture of 
$\cal OS$ initially, {\it cf.} Section\,1.2, which can be incorporated in a density matrix.  

We conclude that in the framework of CAI quantum-classical hybrids are {\it not} a consistent 
construct. With hindsight, this was to be expected, since classical states are of ontological 
kind, while quantum states are mathematical constructs of epistemological character, according to the 
Cellular Automaton Interpretation of QM \cite{tHooft2014}, see also Ref.\,\cite{Rovelli2015}.    

Perhaps, the longstanding difficulties faced by all   attempts to find a satisfactory description of quantum-classical composites in quantum theory, {\it cf.} Section\,1.1 and see, {\it e.g.},   
Refs.\,\cite{Elzehybrid,Salcedo,Elze2012,Diosi2012,Diosi2014} with further references, have indicated already that such attempts are fundamentally flawed. It is interesting that CAI reveals this in a very straightforward way, as we have seen. 

\section{Conclusions} 
We have reviewed the concept of and interest in 
Quantum-Classical Hybrids (QCH), {\it i.e.} composites of at least one classical and one quantum mechanical part, followed by a sketch of the Cellular Automaton Interpretation (CAI) of quantum mechanics. 

We elaborated some aspects of the latter in models of chains consisting of two-state Ising spins, which interact by transpositions of spins that is by permutations of the states of a chain. The latter are  considered as {\it ontological states} (${\cal OS}$), existing ``out there'' in  this model Universe. 
The resulting deterministic dynamics is generated 
by fine-tuned self-adjoint Hamiltonian operators, 
much like in QM. 

Perturbations of such Hamiltonians, generally, necessitate an enlarged state space for the ensuing  approximate description of the model. It is the Hilbert space that admits superposition states of the  ${\cal OS}$ as an {\it epistemic} mathematical construct. This is where {\it quantum states} enter. 

Based on these observations, the description of QCH 
can be embedded in CAI, as shown in Section\,4. Then, the essential interaction between classical and quantum mechanical components can be 
introduced, as we have illustrated by a generic example, and its effect on the quantum-classical composite structure be analyzed. 

This reveals that QCH of interacting quantum mechanical and ontological degrees of freedom are 
{\it inconsistent}: Unavoidably, interactions, 
$\hat {\cal I}:\{{\cal OS}\}\times\{{\cal OS}\}' \rightarrow\{{\cal OS}\}\otimes\{{\cal OS}\}'$, which bilinearly map ${\cal OS}$ of both chains to 
tensor product states in a resulting Hilbert space, 
transfer quantumness in the form of (entangled)   superpositions also to the assumed classical sector of a QCH. Thus, the hybrid feature is lost through interactions and the concept of QCH looses its meaning 
within CAI, which is intended to address the hypothesis that the Universe is fundamentally of  discrete deterministic character. 
	
Several concluding remarks are in order here: 

Since we have demonstrated {\it inconsistency of 
quantum-classical hybrids} from the perspective of the Cellular Automaton Interpretation of quantum mechanics, this should, of course, not detract from their practical importance in approximation schemes  for otherwise still impossible studies of complex objects, {\it cf.} Section\,1.1. 

It may be obvious that cellular automaton models 
related to CAI  
fall into the class of {\it superdeterministic} 
dynamical models, which recently have found renewed interest -- especially, in order to show that  
the verdict of Bell's theorem against realistic local  hidden variable theories can be circumvented \cite{tHooft2014,Vervoort1,Vervoort2,Vervoort3}. 

The interaction between two chains that was instrumental for our present argument, of course, has been chosen {\it ad hoc} as an example. It is left for future work to construct models along these lines that 
may approximate physically relevant continuum field 
theories. It would be very interesting to understand 
the relation between such {\it interactions} and  
{\it computational gates}. 

\begin{acknowledgements}
It is a pleasure to thank Louis Vervoort and 
Theo Raptis for correspondence and Ken Konishi for discussions on various related matters.  
The organizers of {\it The Quantum} \& {\it The Gravity 2021} 
are thanked for the kind invitation to 
present this work. 
\end{acknowledgements}

\section*{Conflict of interest and other statements}
The author declares that he has no conflict of interest. -- This work received no external funding and has been performed solely by the author.

\end{document}